%% file: main.tex
\documentclass[structabstract]{aa}  
\pdfoutput=1 

\usepackage{amsmath}

\usepackage{graphicx}
\usepackage{txfonts}
\usepackage{natbib}

\usepackage{color}

\input{new_commands.tex}

\begin{document}
\title{Towards the field binary population:\\ Influence of orbital decay on close binaries}

% \subtitle{}

\author{C. Korntreff\inst{1} \and T. Kaczmarek\inst{2} \and S. Pfalzner\inst{2}}

\institute{
  \inst{1}J\"ulich Supercomputing Centre, Forschungszentrum J\"ulich GmbH, 52425 J\"ulich, Germany\\
  \inst{2}Max-Planck-Institut f\"ur Radioastronomie, Auf dem H\"ugel 69, 53121 Bonn, Germany\\
 \email{t.kaczmarek@mpifr-bonn.mpg.de}
}

\date{}

\input{abstract.tex}
\titlerunning{Orbital decay}
\authorrunning{Korntreff, Kaczmarek \and Pfalzner}
\maketitle

% 
% ________________________________________________________________

\input{introduction}

% __________________________________________________________________

\section{Cluster Setup}

\label{sec:clust-binary-popul}
\input{cluster_and_binary_setup.tex}

% __________________________________________________________________

\section{Gas-induced Orbital Decay}
\label{sec:embedded_phase}
%\subsection{Method}
%\label{sec:embedded_phase_method}
\input{method_decay}

%\subsection{Results}
%\label{sec:embedded_phase_results}
\input{results_decay}
% 
% ______________________________________________________________

%\section{Dynamical Destruction}
%\label{sec:nbody}
%\subsection{Method}
%\label{sec:nbody_method}
%\input{method_nbody}
%\subsection{Results}
%\label{sec:nbody_results}
% \input{results} 
% % 
% % ______________________________________________________________

\section{Comparison with observations}
\label{sec:combined-results}

\input{combined_results}

% % 
% % ______________________________________________________________

%% \section{Discussion}
%% \label{sec:discussion}

%% \input{discussion}
% % 
% % ___________________________________________________________

\section{Conclusions}
\label{sec:conclusions}

\input{conclusions}
% 
% ______________________________________________________________

\begin{acknowledgements}
  We thank the referee for providing constructive comments and help in improving the contents of this paper and P. Kroupa for useful discussions on this topic. The simulations were performed at the J\"ulich Supercomputer Centre, Research Centre J\"ulich within Project HKU14.
\end{acknowledgements}

% Bibliography.
% -------------
\bibliographystyle{aa}
\bibliography{mybib,reference}

% \begin{thebibliography}{}

% \end{thebibliography}

\end{document}

%% file: new_commands.tex
\newcommand {\yr} {\mbox{yr}}

\newcommand {\nbody} {\textsc{\mbox{nbody6}}}
\newcommand {\nbodypp} {\textsc{\mbox{nbody6\raise.4ex\hbox{\tiny++}}}}
\newcommand {\secnbodypp} {\textsc{\mbox{nbody6\raise.4ex\hbox{\scriptsize++}}}}

\newcommand {\Msun} {\mbox{M$_{\odot}$}}

\newcommand {\days} {\mbox{d}}
\newcommand {\Myr} {\mbox{Myr}}

\newcommand {\AU} {\mbox{AU}}
\newcommand {\pc} {\mbox{pc}}

\newcommand {\Rhm} {\mbox{R$_{\textrm{hm}}$}}

\newcommand {\Nbin} {\mbox{$N_b$}}

\newcommand {\Msys} {\mbox{$M_{\mathrm{sys}}$}}

\newcommand {\Nbinnorm} {\mbox{$\mathcal{N}_b$}}
\newcommand {\Nmergnorm} {\mbox{$\mathcal{N}_m$}}

%% file: abstract.tex
\abstract{Surveys of the binary populations in the solar neighbourhood have shown that the periods of G- and M-type stars are
  \textit{log-normally} distributed in the range from $0.1-10^{11}$ days. However, observations of young binary populations in
  various star forming regions suggest a \textit{log-uniform} distribution. Clearly some process(es) must be responsible for this
  change of the period distribution over time. Most stars form in star clusters, so it is here that the(se) process(es) take place.}
{In  dense young clusters two important dynamical processes occur: i) the gas-induced orbital decay of embedded binary systems and ii) the
  destruction of soft binaries in three-body interactions. The emphasis in
  this work is on orbital decay as its influence on the binary distribution in clustered environments has been largely neglected so far.}
{We performed Monte-Carlo simulations of binary populations to model the process of orbital decay due to friction with the
  gas.  In addition, the destruction of soft binaries in young dense star clusters was simulated using nbody modelling of binary populations.}
{It is known that the cluster dynamics destroys the number of wide binaries, but leaves short-period binaries basically undisturbed. 
Here it is demonstrated that this result is also valid for a initially log-uniform period binary distribution. By contrast the 
process of orbital decay significantly reduces the number and changes the properties of short-period binaries, but leaves wide binaries 
largely uneffected. Until now it was unclear whether the short period distiribution of the field is unaltered since its formation. 
It is shown here, that if any alteration took place, then orbital decay is a prime candidate for this task. In combination the dynamics 
of these two processes, convert evan an initial log-uniform distribution  
to a log-normal period distribution. The probability is $94\%$ that the evolved period distribution and the observed period distribution have been sampled from the same parent distribution.} 
{Our results provide a new picture for the development of the field binary population: Binaries can be formed as a result of the star-formation process in star clusters with periods that are sampled from the log-uniform distribution. As the cluster
  evolves, short-period binaries are merged to single stars by the gas-induced orbital decay while the dynamical evolution in the
  cluster destroys wide binaries. The combination of these two  equally important processes reshapes a initial log-uniform period
  distribution to the log-normal period distribution, that is observed in the field.}

% 5 {} token are mandatory

%%% Local Variables: 
%%% mode: latex
%%% TeX-master: "main"
%%% End: 

%% file: introduction.tex
\section{Introduction}
\label{sec:introduction}

%
% |-------+--------+--------+--------+-----------+-------------------|                                                                      
% | type  | single | double | triple | quadruple | calculated double |                                                                      
% |-------+--------+--------+--------+-----------+-------------------|                                                                      
% | DM91  |     50 |     41 |      7 |         2 |                61 |                                                                      
% | Rag10 |     56 |     33 |      8 |         3 |                58 |                                                                      
% | FM92  |     58 |     33 |      7 |         1 |                50 |                                                                      
% |-------+--------+--------+--------+-----------+-------------------|
% 

The most important observations that shaped our current picture of the binary field population were already performed in the
early 1990's. \citet{1991A&A...248..485D} determined the binary frequency of G-type stars in the solar neighbourhood to be
about $61\%$ and the period distribution to follow a log-normal distribution, 

\begin{equation}
\label{eq:dm91_period_distribution}
f(\log P) = \mathrm{C} \exp\left\{ \frac{ -( \log P - \overline{\log P})^2 }{ 2\sigma_{\log P}^2 } \right\}, 
\end{equation}

over the period range $\sim 10^{-1}- 10^{11}\days$ where P is the period in days ($\overline{\log P} = 4.8 \equiv 172 \yr$, $\sigma_{\log P}=2.3$) and $\mathrm{C}$ a normalisation
constant. A year later \citet{1992ApJ...396..178F} analysed the binary properties of M-dwarfs in the separation range
$0.04-10^4\AU$ corresponding to periods in the range $\sim 10^{-2} - 10^{6}\yr$ and found that the period distribution of
M-dwarfs in the solar neighbourhood is also log-normally distributed with a peak between $9\yr$ and $270\yr$ - nearly identical
to the findings for the G-dwarfs. Although the observed binary frequency of M-dwarfs with about $33\%$ is lower than that of
G-dwarfs, the properties of the G and M-dwarf binary populations seem very
similar. More recent observations by \citet{2010ApJS..190....1R} confirm this
picture. They determined the binary properties of nearby ($d < 25\pc$)
solar-type stars ($\sim \mathrm{F}6 - \mathrm{K}3$) and find the fraction of
binary stars to be $58\%$ with a log-normal period distribution
($\overline{\log P} = 5.03$ and $\sigma_{\log P}=2.28$). 

With an average age of a few Gyrs the field population constitutes an  dynamically evolved state. Therefore, dynamical processes
have most likely changed the binary properties since the formation of these stars and the current
properties differ significantly from the primordial state. In order to understand the binary formation process it is not
sufficient to know the properties of the (old) field population but  those of the primordial binary population.

\citet{2008AJ....135.2526C} determined the binary properties of the very young populations in Taurus, Ophiuchus and the Orion
star-forming regions (excluding the  much denser Orion Nebula Cluster). They found that their observed period distribution can be fitted by a
log-uniform distribution, differing significantly from the log-normal distribution in the field (see also \citet{2007ApJ...662..413K}). In these sparse young star forming regions it is
improbable that dynamical evolution altered the period distributions in the short period since the stars were formed \citep[see
e.g.][]{2003MNRAS.346..343K}. So it can be assumed that their properties match the initial conditions in general. 

In denser regions indications for an originally log-uniform distribution are found, too. For example, HST observations by
\citet{2007AJ....134.2272R} of binaries in the Orion Nebula Cluster (ONC) demonstrate, that the semi-major axis distribution
deviates from the log-normal distribution and is closer to a log-uniform distribution. So it seems that older binary populations
have a \textit{log-normal} period distribution while the primordial distribution is likely to be \textit{log-uniform}.

From the theoretical side the binary frequency is the best studied binary property
\citep[e.g.][]{2001ApJ...555..945K,2003MNRAS.346..343K,2001MNRAS.321..699K,2007A&A...475..875P,2011arXiv1106.5050M,2011arXiv1109.2896M}. It
seems that also  the evolution of the binary frequency in the ONC  depends on
the initial binary frequency, the evolution of the binary population does not
\citep{2011A&A...528A.144K}. However, the binary population is not only
described by the binary frequency but also by the period, mass-ration and
eccentricity distribution. 

Starting with the work by \citet{1975MNRAS.173..729H}, it has been realised
that binaries become dynamically destructed via three-
and four-body interactions. Generally wide binaries are more affected by
dynamical destruction than close ones. The existence of binaries wider than
$10^{3}\AU$ in the field still poses an open question regarding their origin
\citep{2009MNRAS.397.1577P}. 

\begin{figure}
  \centering
  \includegraphics[keepaspectratio,width=0.9\linewidth]{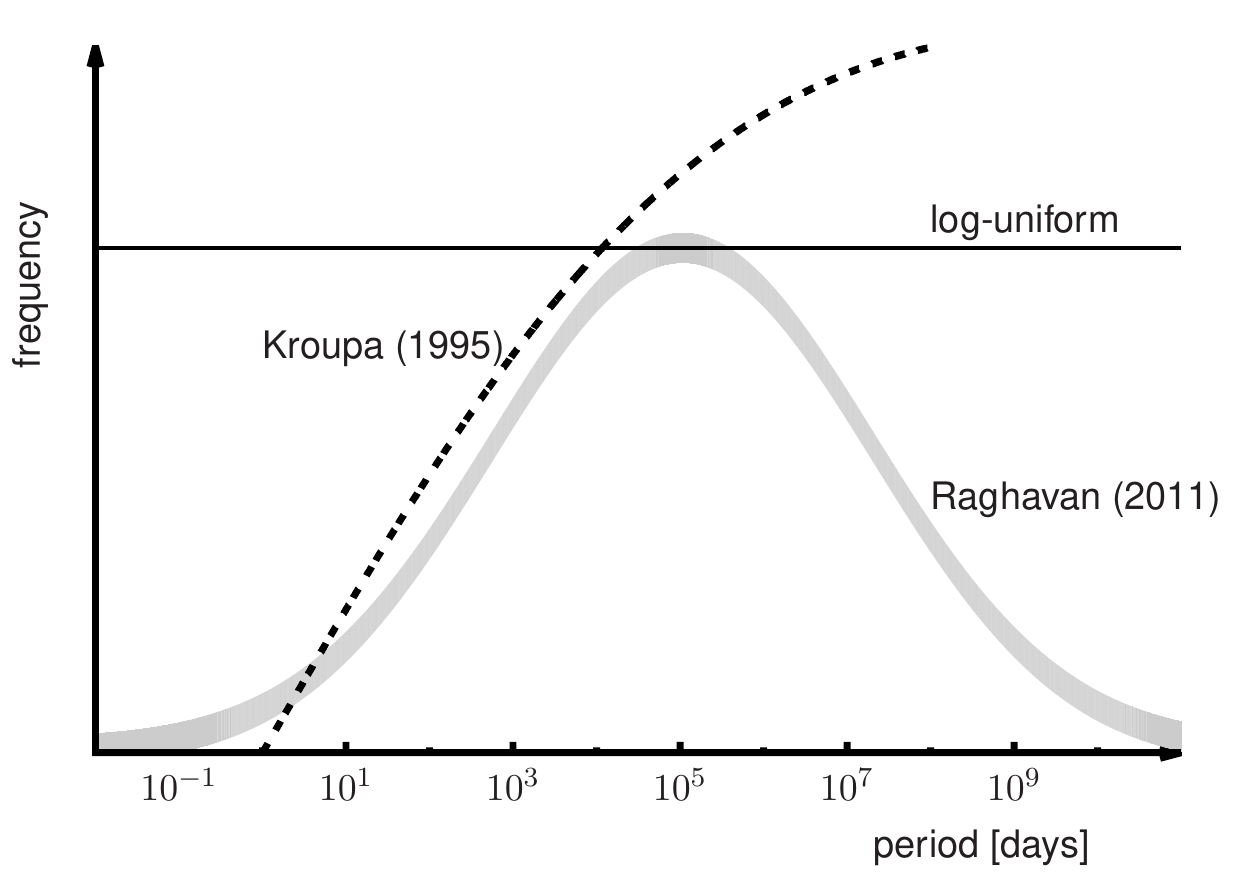}
  \caption{Schematic picture of possible initial period distributions. The thick grey curve displays the log-normal fit to the
    data of \citet{2010ApJS..190....1R}, the thick dashed line the \citet{1995MNRAS.277.1491K} and the solid line a log-uniform
    period distribution, that is observed in many young star forming regions (see text for references).}
  \label{fig:possible_period_distributions}
\end{figure}

Performing N-body simulations \citet{1995MNRAS.277.1491K,1995MNRAS.277.1522K}
showed, that to reproduce the observed log-normal distribution of the field,
the initial number of wide binaries has to be significantly higher than 
observed, if all binaries are exposed to dynamical evolution  in a star cluster. 
This rising distribution was obtained by inverse dynamical population
synthesis, inverting the effects of dynamical destruction on the period
distribution (see dashed line in Fig.~\ref{fig:possible_period_distributions}). 

\citet{2009MNRAS.397.1577P} finds that binaries with periods exceeding
$10^{3}\AU$ cannot survive in these clusters he investigated. However, binaries
with semi-major axes exceeding $10^{3}\AU$ are observed in the
field. \citet{2010MNRAS.404.1835K} suggest, that these observed wide binaries
could form during the cluster dissolution. Another approach would be a
different cluster type.
%% for example as in \citet{1995MNRAS.277.1491K}. 

Here we investigate how the initial period distribution would evolve in its
natal environment considering the two different dynamical processes that
inevitably affect binaries in a cluster: i) gas-induced orbital decay and ii)
dynamical destruction of binaries in encounters. 

So far the influence of orbital decay has been only investigated for isolated
binary systems \citet{2010MNRAS.402.1758S}.  Orbital decay takes place in the
earliest stages of star formation, when the stars are still embedded in the
gas. Here the interaction of binaries with the gas leads to the 
excitation of waves in the surrounding gas. The energy transfer from the 
binary to the gas leads to orbital decay \citep{2010MNRAS.402.1758S}. 
Here we investigated how orbital decay changes the period distribution in a 
cluster environment. 

This paper is structured in the following way: in
Section~\ref{sec:clust-binary-popul} the cluster setup is
explained. Section~\ref{sec:embedded_phase} describes the orbital decay
process in the gas embedded phase. We demonstrate in
Section~\ref{sec:combined-results} that a combination of the orbital decay
with the dynamical destruction \citep{2011A&A...528A.144K} naturally
transforms a log-normal birth period distribution into a log-uniform period
distribution. Our conclusions mark the last section. 

%%% Local Variables: 
%%% mode: latex
%%% TeX-master: "main"
%%% End: 

%% file: cluster_and_binary_setup.tex
The processes of orbital decay and dynamical destruction take inevitably
  place as soon as the binary population forms in a dense star cluster environment.
  Both result from interactions between stars in a dense environment. As most stars 
form in cluster environments \citep{2003ARA&A..41...57L}, it can be anticipated that 
the field binary population we observe today was - at least to same extend - shaped by these processes. 

In this study we chose the Orion Nebula Cluster (ONC) as model cluster. It is
probably a typical star forming region and has the advantage that many
parameters both of the cluster structure and the binary population are well
observed
\citep[e.g.][]{1988AJ.....95.1755J,1994ApJ...421..517P,1997AJ....113.1733H,1998ApJ...492..540H,2009AJ....137..367O,1999NewA....4..531P,2006A&A...458..461K,2007AJ....134.2272R}.
The radial density profile declines in the outer parts as $\rho_{\mathrm{present}} \propto r^{-2}$
\citep{1988AJ.....95.1755J,1997AJ....113.1733H} and is much flatter ($\rho_{\mathrm{core}} \propto r^{-0.5}$) within
$R_{\mathrm{core}}\approx 0.2\pc$. The dynamical evolution of the cluster significantly changes the stellar density profile during
the first Myr, the estimated age of the ONC \citep{1997AJ....113.1733H}. Hence we adopted the following initial density profile
\begin{align}
\label{eq:radial_density_profile__initial}
\rho_{\text{star}}(r) \propto \left\{\begin{array}{ccc} 
 (r/R_{\text{core}})^{-2.3}, & 0 < r \leq R_{\text{core}}\\ 
 (r/R_{\text{core}})^{-2.0}, & R_{\text{core}} <r \leq R   \\
0, & R < r \leq \infty .
\end{array} \right.
\end{align}
Here we choose  $R_{\text{core}}=0.2$ pc and a cluster size of
$R=2.5$ pc to model the ONC. The used model starts from a situation were all
stars are already formed. In dynamical equilibrium the initial radial density
profile (Eq.~\ref{eq:radial_density_profile__initial}) evolves towards the
currently observed one over a time span of about 1Myr
\citep{olczak_stellar_2010}. However, other initial conditions could result in
the same profile \citep{2000NewA....4..615K,2009ApJ...700L..99A}.

To mimic the often observed mass-segregation in young dense clusters, the
most massive binary system is initially placed at the cluster centre and the
three next most massive stars at random positions within a sphere of radius
$R=0.6\,\Rhm$ around the cluster centre, where \Rhm\ is the cluster half-mass
radius. In this we follow the suggestions by \citet{1998MNRAS.295..691B}, who
found that the observed mass segregation \citep{1998ApJ...492..540H} of the
ONC is unlikely to be the result of the dynamical evolution of the cluster but
has to be primordial, if the cluster is initially in virial equilibrium. Note, that if the cluster would be in a subvirial state, probably no primordial mass segregation would be required \citep{2009ApJ...700L..99A,2011arXiv1108.2446O}.

All stellar systems are initially created as binaries with
system masses sampled from the initial mass function (IMF) suggested by
\citet[][Eqs. 19 \& 20]{kroupa_initial_2007}. We limited our system mass
$\Msys = M_{s1} + M_{s2}$ to $[0.08\,\Msun,50\,\Msun]$, where $M_{s1}$ and $M_{s2}$ are the masses of the
stars constituting a binary system. The upper mass limit 
corresponds to the mass of the most massive binary system of the ONC. The lower limit is the hydrogen burning limit, and thus the margin between sub-stellar and
stellar objects.

Ideally we would like to use observed initial properties of binaries in very young embedded clusters. We will see that embedded clusters younger
  than 100,000 yr would be ideal, because at such a young age the binary population  would be close to its primordial state. However, it is very difficult 
  to observe clusters at such a young age. Observations of binaries in dense embedded clusters face a number of
  observational difficulties like extinction, crowding, resolution etc.. In addition, the number of stars would be very low, making statistical statements impossible. As their is an age spread
  in forming clusters (1-3 Myr) it would be a possibility to use just the youngest stars in a forming cluster to deduce the primordial state. However, 
  age determination for such young stars is problematic. Alternatively, one can assume that low density clusters are relatively
    unaffected by these processes described here and are therefore close to their primordial state. Observations of the primordial state not
    being available, we use the work by \citet{kouwenhoven_primordial_2007}
    based on observations of the low-density Scorpius OBII cluster. Choosing the semi-major axis and mass ratio distribution from an
OB association with an age of (5-20) Myrs \citep{kouwenhoven_primordial_2007},
as Scorpius OBII, seems at first glance unusual, as at that age the binary
population might have already been processed. 
However, the low stellar density of Scorpius OBII combined with the lack of
surrounding gas speaks for the cluster having neither been processed by
orbital decay nor by dynamical destruction, as mentioned before.

The primordial semi-major axis distribution was chosen to be log-uniform
$f_a(a) \propto d\mathrm{N} / da \propto a^{-1}$
 with $a \in [0.02\,\AU,10000\,\AU]$. As primordial mass ratio distribution
we started with $f_q(q) \propto q^{0.4}$ and $q=M_{s2}/M_{s1}
\in [0,1]$. 

Another justification for our choice of the initial semi-major axis distribution, is that
the log-uniform distribution is a simple straight forward distribution, that
requires strong processing in order to obtain the log-normal field
distribution. This distribution was used here to test,
if even this extreme assumption leads to the log-normal distribution observed
in the field. This does not rule out, that one obtains a similar result as
described in the following with different initial conditions, containing
fewer wide and/or close binaries.

In the following the above described model will be used to study the effect of
the gas-induced orbital decay of binaries in star clusters.

%%% Local Variables: 
%%% mode: latex
%%% TeX-master: "main"
%%% End: 

%% file: method_decay.tex
In the early embedded phases clusters consist of the already formed stars and a large gas component from which further stars
potentially form. These stars keep being embedded in their natal gas cloud until it is removed by strong winds, radiation of
the massive stars and supernova explosions. During this embedded phase, stars and binary systems experience dynamical friction with the ambient gas.
 Binary systems induce spiral waves in the gas leading to energy and angular momentum loss resulting in a shrinking of the binary orbit
and possibly merging of the two stars. \citet{2010MNRAS.402.1758S} derived an analytic expression for the temporal development of
the separation $a_{tot}$ for a single isolated binary system on a circular orbit. The separation diminishes due to gas-induced
orbital decay with time as
\begin{align}
\label{orbital_decay}
a_{tot} = a_0 \left(1- \frac{t}{t_c}\right),
\end{align}
where $a_0$ is the initial binary separation and $t_c$ the so-called coalescence time, which is given by
\begin{align}
\label{orbital_decay2}
t_c = \frac{15}{32 \pi} \frac{(1+q)^2}{q}\frac{c_s^5}{\rho_0}\frac{a_0}{G^3 M_{sys}^2} ,
\end{align}
where $G$ is the gravitational constant, $c_s$ the sound speed, $q$ the mass ratio and $\Msys$ the system
mass of the binary system.

\citet{2010MNRAS.402.1758S} assumed in his derivation that gas interactions close to the stars themselves can be
neglected. The binary system, represented by an oscillating gravitational potential, torques the nearby gas and produces outgoing
acoustic waves. These waves transport angular momentum from the binary to the surrounding gas. Thus the orbit of the binary decays
as long as the gas density is high enough. Eq.~\ref{orbital_decay} is only valid for radial distances
\begin{align}
\label{r_in_again}
  a_{tot} < r^{\text{max}} =\frac{G\,\Msys^\text{{min}}}{2\,c_s^2} =
  111\,\AU \left(\frac{\Msys^{\text{min}}}{M_{\odot}}\right)
  \left(\frac{c_s}{2\, \text{km s}^{-1}}\right)^{-2} ,
\end{align}
between the two stars constituting the binary system.

To analyse the effect of gas-induced orbital decay of binary systems on the period distribution, we applied
Eq.~\ref{orbital_decay} to the entire population of a young star cluster, for the example of the ONC. Starting from a cluster with
initial conditions as given in Sec.~\ref{sec:clust-binary-popul}, the temporal development of the binary population is followed
for different gas density models.

Only if a binary system fulfills the restriction given by Eq.~\ref{r_in_again} the orbital decay is calculated, otherwise the
orbit remains unaltered. This means, for example, for a system mass of $0.08\,\Msun$, $1\,\Msun$ and $50\,\Msun$ the orbital decay
will only be calculated for binary systems with a separation $r < 8.88\,\AU$, $r < 111\,\AU$ and $r < 5550\,\AU$. Like in
Stahler's work modelling the orbital decay the simplification of treating all orbits as circular has been adopted. Future work
should include elliptical orbits as well.

For the gas-induced decay, the gas density distribution in the cluster is of vital importance.  Here we assume an isothermal gas
density which follows the stellar density distribution.  To prevent the distribution from diverging at the centre, the density is
kept constant at a value $\rho_{\text{max}} $ inside the cluster core radius $R_{\text{core}} = 0.2\pc$.  Outside this area the
density decreases isothermally. Thus the isothermal gas density distribution can be described by the following equation

\begin{align}
\label{eq:densitydistribution}
  \rho_{\textrm{gas}}(r) = \rho_{\text{max}}\left\{\begin{array}{cc}
  1, & r < R_{\text{core}} \\ 
  (R_{\text{core}}/r)^2, & R_{\text{core}} < r < R_{\text{cluster}}
  \end{array}\right.
\end{align}

%with $R_{\text{cluster}} = 2.5\,\pc $

For an isothermal density distribution $\rho_{\textrm{gas}}(r) = c_s^2 /2 \pi \, G \, r^2$ \citep[e.g.][Eq. 4.123]{binney_galactic_1987} the
sound speed is given by
\begin{align}
c_s &= \sqrt{2 \pi \, G \,\rho_{\textrm{gas}}(r) \, r^2} .
\end{align} 

Although the density distribution (Eq.~\ref{eq:densitydistribution}) is not
isothermal over the whole parameter range, we approximate the sound speed as
\begin{align}
\label{soundspeed}
 c_s &= \sqrt{2 \pi \, G \,\rho_{\text{max}} \, r_{\text{core}}^2} .
\end{align}

The resulting sound speeds for distributions  with $\rho_{max} \in [10^4, 10^6] cm^{-3}$ lie between the sound speed of an ideal gas at 10K, $c_s
  = 0.2 km/s$, and the observed value in infrared dark clouds \citep{sridharan_high-mass_2005} $c_s = [1.0-2.2] km/s$. A even better approach would be
  a time dependent density $\rho(r,t)$ and sound speed $c_s(r,t)$, which hasn't been observed yet. Deviations of the sound speed manifest themselves in the coalescence time (Eq.~\ref{orbital_decay2}).

If the separation becomes smaller than the radius of both stars $a_{\text{tot}} < R_{s1} +
R_{s2}$ (assuming a main-sequence mass-radius relationship \citep[p. 110]{binney_galactic_1998}),
\begin{align*}
\left ( \frac{R_{s1/2}}{R_{\odot}} \right) \approx \left ( \frac{M_{s1/2}}{\Msun}\right) ^{0.7},
\end{align*}
a binary system is treated as a 'merged' system. This is clearly a strong simplification, because the merging of stars is a much
more complex process that depends on a variety of parameters such as the presence of a surrounding disc, the eccentricity of the
orbit, mass transfer and the conditions of the molecular cloud in which the stars are embedded. However, including all these
processes goes beyond the aim of the current paper. Here, all processes at these small distances are excluded under the assumption
that these two stars might possibly merge at such small distances.
 
%%% Local Variables: 
%%% mode: latex
%%% TeX-master: "main"
%%% End: 

%% file: results_decay.tex
First it was analysed how the binary population parameters change due to orbital decay in general. Therefore, a series of
simulations with three different gas densities in the range of [$10^4\text{cm}^{-3}$, $ 10^6\text{cm}^{-3}$] were performed
corresponding to moderate gas densities observed in star forming regions \citep{padmanabhan_theoretical_2001}.  For the three
densities $ 10^4\text{cm}^{-3}$, $ 10^5\text{cm}^{-3}$ and $ 10^6\text{cm}^{-3}$ the sound speeds are 0.15 km/s, 0.49 km/s and
1.54 km/s, respectively (Eq.~\ref{soundspeed}).  The results summarised in Table~\ref{tab:merger}, show the merger rates
for all combinations of gas densities and sound speeds. As to be expected, the percentage of possibly merged binary systems after
$1\,\Myr$ (Table~\ref{tab:merger}) depends on the maximum density of the isothermal density distribution
(Eq.~\ref{eq:densitydistribution}). The value of the overall reduction of binaries could vary for a constant sound speed by a
factor of two to four if the density would change by one order of magnitude. Keeping the density constant, one can see, that the
percentage of merged systems is sensitive to the choice of sound speed.

Additionally, the degree of orbital decay depends on the gas density distribution. Here only
the case of a temporary constant and spatially isothermal gas density distribution has been described as this fits best the
observed period distribution and seems a realistic assumption for the here studied case of an ONC-type cluster.
 
In the following an isothermal gas density distribution with $\rho_\text{max} = 10^5\text{cm}^{-3}$ is used for the example of an
ONC-like cluster.

\begin{table}[htbp]
  \centering
  \begin{tabular}{c|ccc}
    \hline
    sound speed & 0.15 km/s & 0.49 km/s & 1.54 km/s \\    
    maximum density & & &  \\
    \hline
    $10^4\text{cm}^{-3}$ & 37.4\% &  2.8\% &  0.1\%  \\
    $10^5\text{cm}^{-3}$ & 54.9\% &  12.0\% &  0.2\%  \\
    $10^6\text{cm}^{-3}$ & 69.2\% &  28.7\% &  1.2\%  \\
     \hline
   \end{tabular}
  \caption{Percentage of merged binary systems for different sound speeds and maximum densities after 1 Myr.}
  \label{tab:merger}
\end{table} 

\begin{figure}
  \centering
  \includegraphics[keepaspectratio,width=0.9\linewidth]{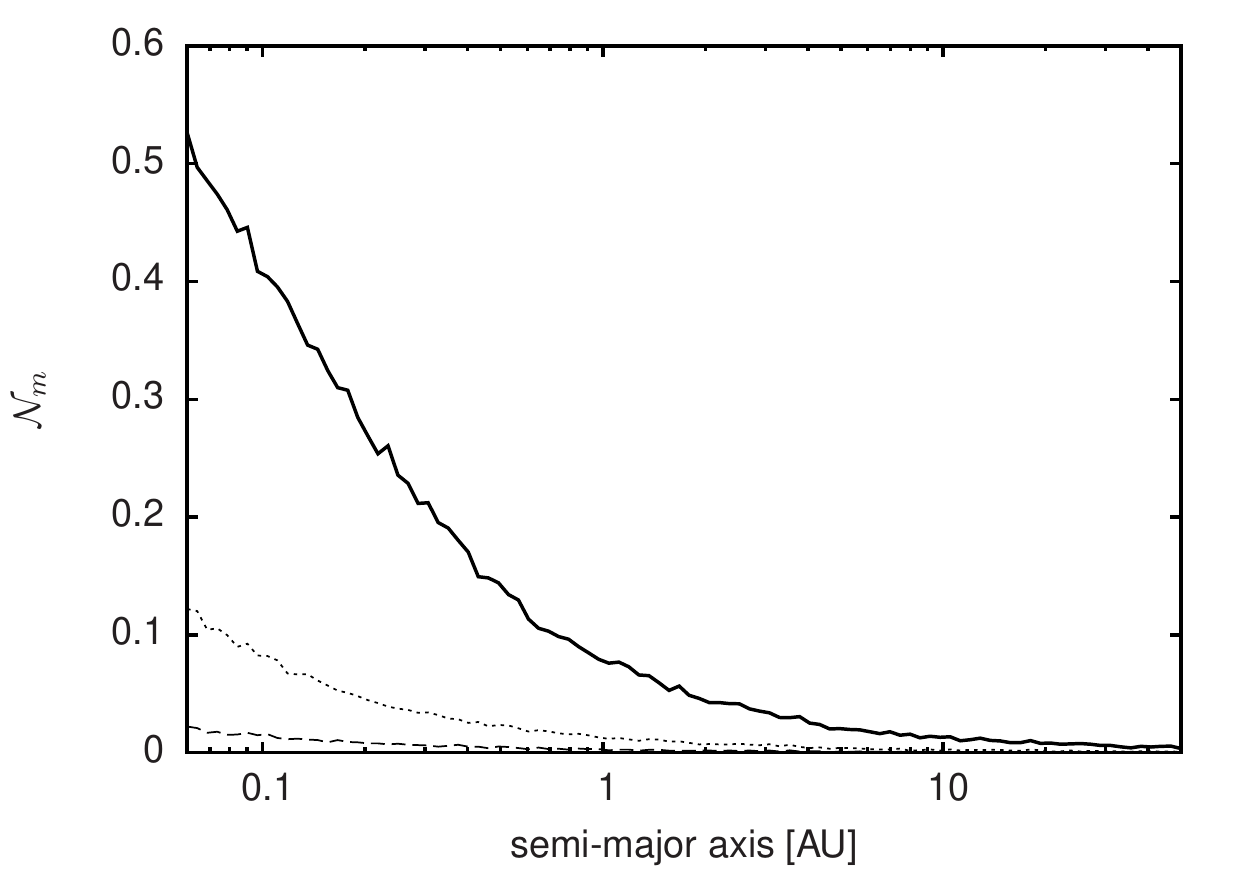}
  \caption{The number of merged binaries relative to the initial number of binaries after $0.01\,\Myr$ (dashed), $0.1\,\Myr$
    (dotted) and $1\,\Myr$ (solid) for a binary distribution embedded in an $r^{-2}$ gas density distribution with
    $\rho_{max}=10^5~\text{cm}^{-3}$ and a sound speed of $0.49$~km/s.}
    \label{fig:zero_semi_major_axis}
\end{figure}

In Fig.~\ref{fig:zero_semi_major_axis} the number of merged binaries $N_m$ relative to the initial number of binaries
$N_{\text{tot}}$, $\Nmergnorm = N_m / N_{\text{tot}}$ is plotted as a function of the initial semi-major axis of the merged binary
systems. The binaries with small semi-major axis merge to a higher degree than wider binaries.  Whereas binaries with initial
semi-major axis of $10\,\AU$ rarely merge, $40\%$ of all binaries with initial semi-major axis of $0.1\,\AU$ coalesce after
$1~\Myr$ (solid line).

\begin{figure}
  \centering
  \includegraphics[keepaspectratio,width=0.9\linewidth]{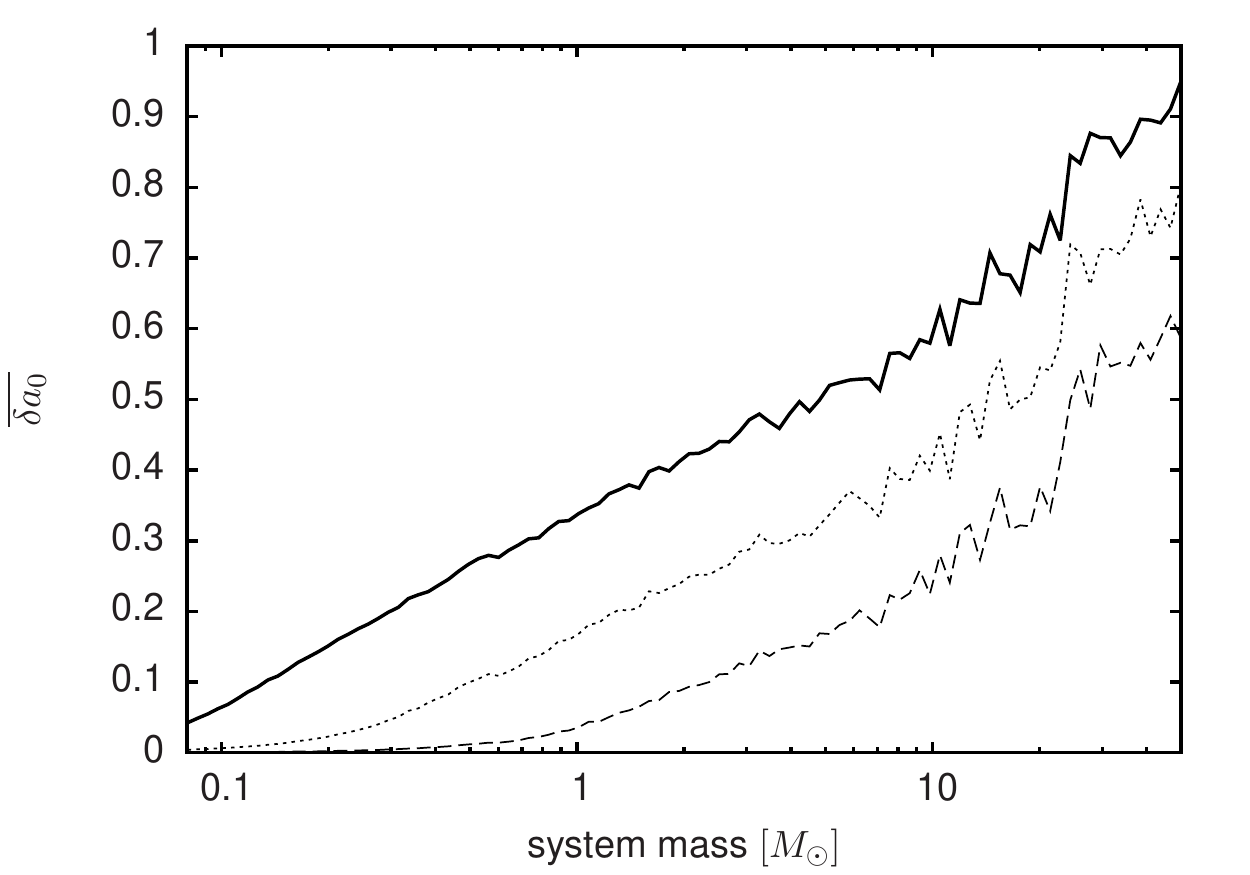}
  \caption{Mean relative semi-major axis change as a function of the system mass for a binary distribution embedded in an $r^{-2}$
    gas density distribution with $\rho_{max}=10^5~\text{cm}^{-3}$ and a sound speed of $0.49$~km/s after $0.01\,\Myr$ (dashed),
    $0.1\,\Myr$ (dotted) and $1\,\Myr$ (solid).}
  \label{fig:relchange_total_mass}
\end{figure}

The resulting semi-major axis change depends as well strongly on the system mass \Msys.  Figure~\ref{fig:relchange_total_mass}
shows the mean relative semi-major axis change $\overline{\delta a_{0}} = [a(0)-a(t)]/a(0) $ as a function of the system
mass. This illustrates that the semi-major axis of binaries with a high system mass shrinks faster and to a larger degree than
the semi-major axis of binaries of a lower system mass. The shorter coalescence time for massive systems (see
Eq.~\ref{orbital_decay2}) results from the increased angular momentum and energy loss of high-mass binaries due to dynamical
friction (Eq. 51 in \citet{2010MNRAS.402.1758S}). The average separation between binary systems with $M_{sys} > 10\, \Msun$
dwindles to less than half its initial value after $1\,\Myr$. Additionally, the location of the binary system inside the cluster influences the
  coalescence time. For a maximum density of $10^5 cm^{-3}$ in the cluster core, the density drops to $6.4\, 10^2\, cm^{-3}$ at 2.5 pc. Therefore the
  coalescence time of a binary system with $M_{sys} = 10\, \Msun$ drops from 0.23 Myr in the cluster center to 35.95 Myr at 2.5 pc. Considering this,
  the effect of orbital decay could even boost the mass segregation of the cluster.

\begin{figure}
  \centering
  \includegraphics[keepaspectratio,width=0.9\linewidth]{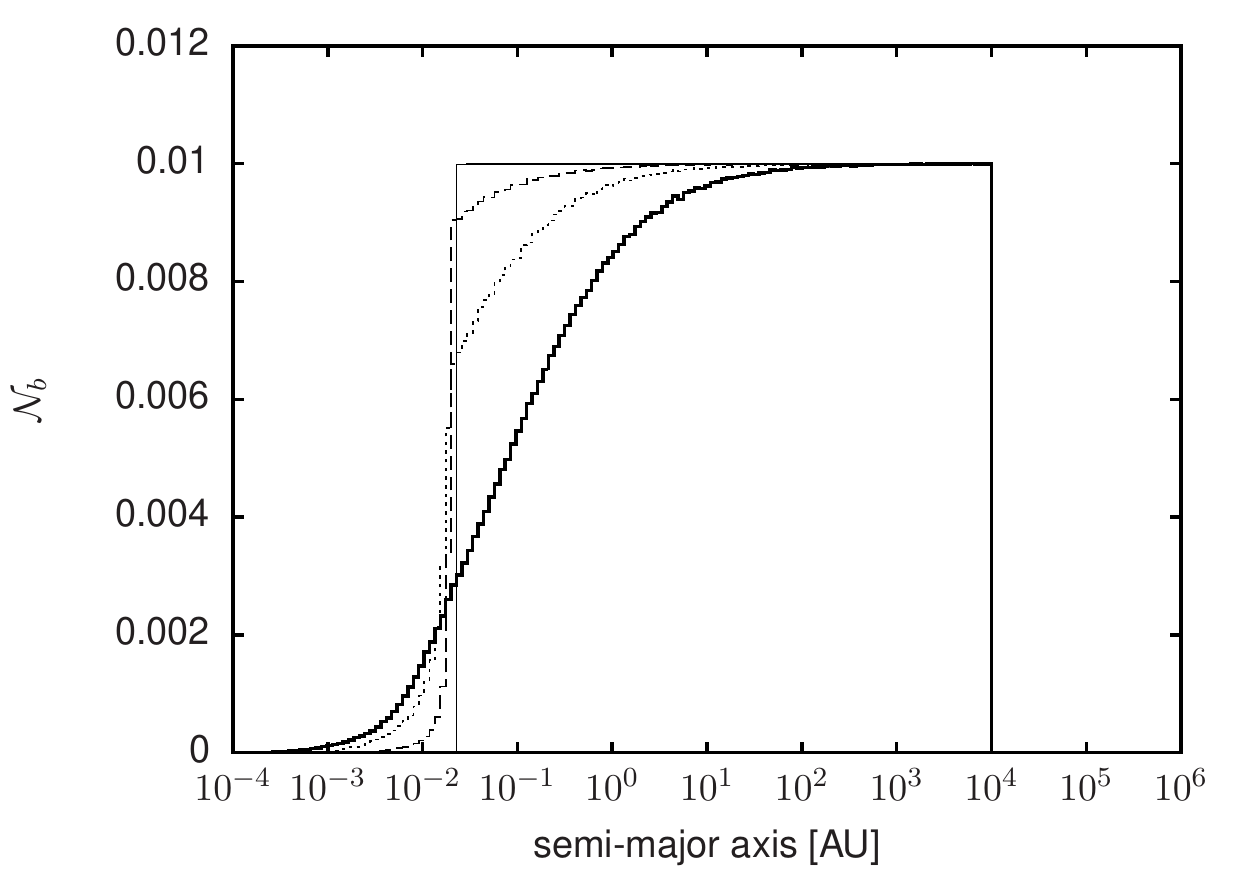}
  \caption{Chosen birth semi-major axis distribution (thin solid line) and effect of the gas induced orbital decay on a semi-major axis
    distributions embedded in an $r^{-2}$ gas density distribution with $\rho_{max}=10^5~\text{cm}^{-3}$ and a sound speed of
    $0.49$~km/s after $0.01~\Myr$ (dashed), $0.1~\Myr$ (dotted) and $1~\Myr$ (thick solid).}
  \label{fig:period_0_0.1_1}
\end{figure}

The overall effect of the gas-induced orbital decay on the period distribution of the binaries in the cluster is to reshape it by
pushing binaries to tighter orbits which can even lead to the merging of a binary. This can be seen in
Fig.~\ref{fig:period_0_0.1_1}, which shows the relative number of binary systems $\Nbinnorm = \Nbin(t) / \Nbin(0)$ as a function
of the semi-major axis after $0~\Myr$ (thin solid line), after $0.01~\Myr$ (dashed line), $0.1~\Myr$ (dotted line) and $1~\Myr$
(thick solid line) for a binary distribution embedded in a $r^{-2}$ gas density distribution with $\rho_{max}=10^5~\text{cm}^{-3}$
and a sound speed of $0.49$~km/s. Because the orbital decay acts faster for tighter binaries (see. Eq. \ref{orbital_decay2}),
binaries in the left part of the period semi-major axis distribution are altered first leading to a depopulation of binaries with
semi-major axes between $0.02\AU$ and $1\AU$ and the formation of a tail for semi-major axis less than $0.02\AU$ after
$0.01\Myr$. As time goes on, binaries with even larger orbits are affected until the orbital decay process stops when the gas is
expelled from the cluster. At this point the process of orbital decay alone is responsible for the binary frequency in the 
cluster to drop from it's initial value of $100\%$ to $88\%$ through the merging of very tight binaries.

%%% Local Variables: 
%%% mode: latex
%%% TeX-master: "main"
%%% End: 

%% file: combined_results.tex
In the following we compare above described results with observations. In order to do so, we converted our semi-major axis
distributions into period distributions and re-binned our data according to \citet{2010ApJS..190....1R}.  Their
observations were limited to F6 to K3 dwarf stars which roughly corresponds to binary systems with primary masses
$M_{s1} \in [0.5~\Msun ,1.5~\Msun ]$.

\begin{figure}
  \centering
  \includegraphics[keepaspectratio,width=0.9\linewidth]{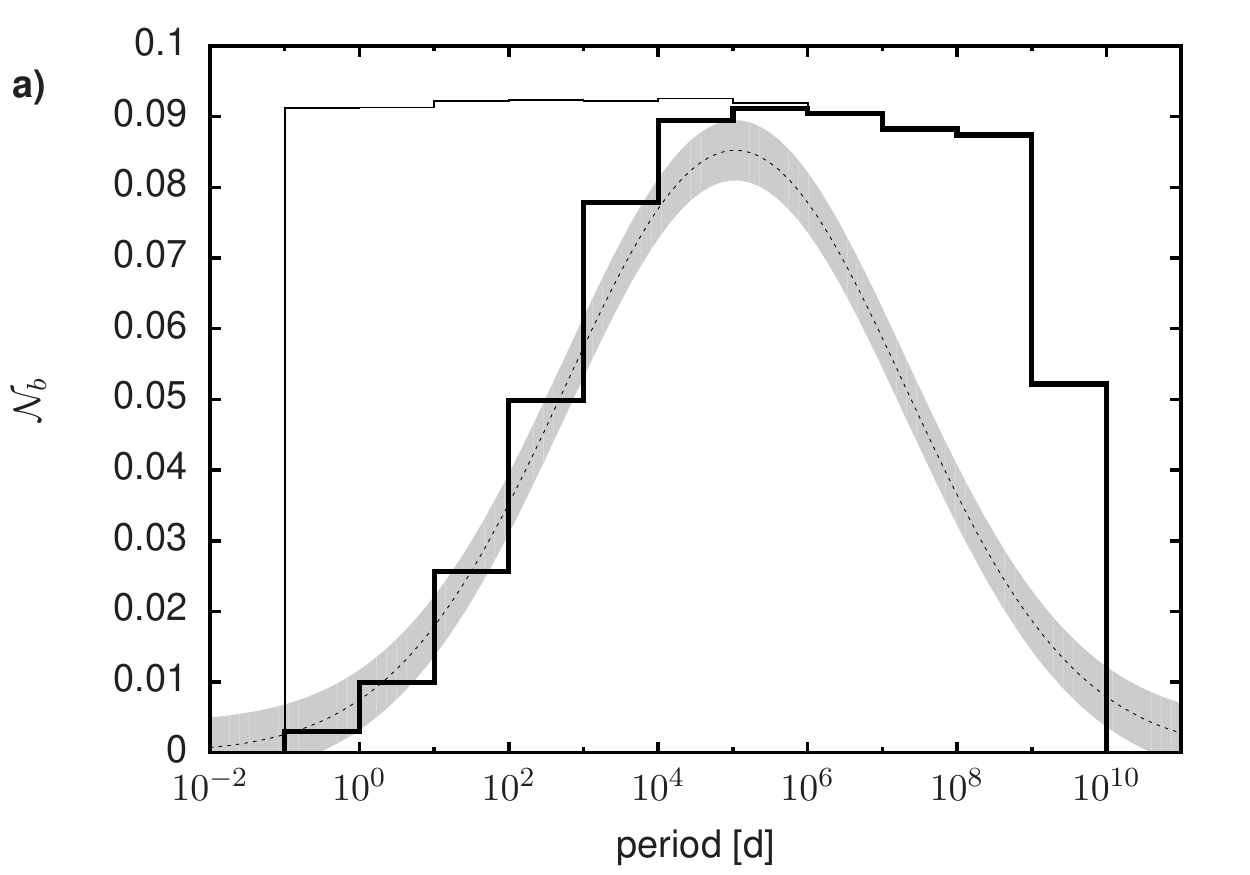}
  \includegraphics[keepaspectratio,width=0.9\linewidth]{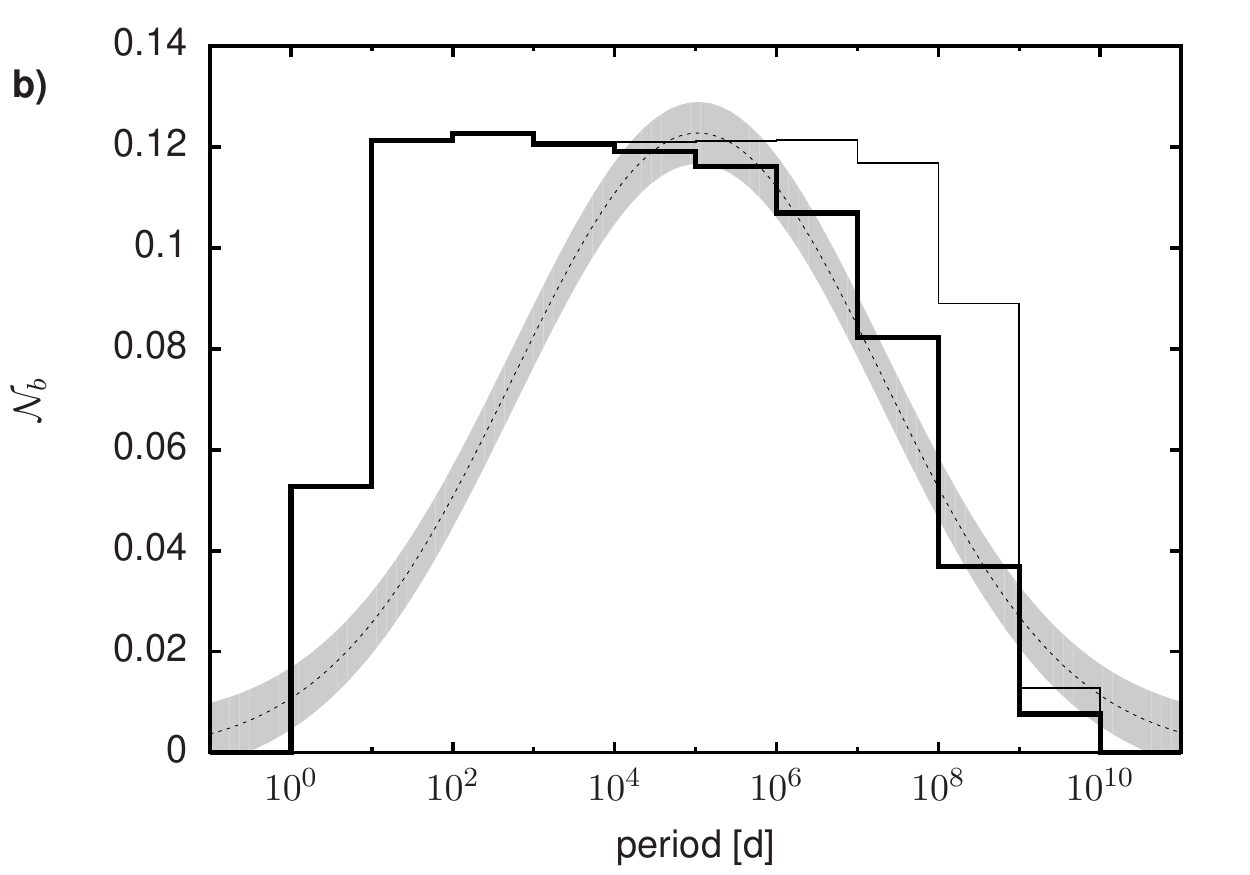}
  \caption{a) Chosen birth period distribution (thin solid line) and effect of the gas induced orbital decay on a period distributions
    embedded in an $r^{-2}$ gas density distribution with $\rho_{max}=10^5~\text{cm}^{-3}$ and a sound speed of $0.49$~km/s after
    $1~\Myr$ (thick solid line) compared with the Gaussian fit (dotted line) to the observations by \citet{2010ApJS..190....1R}. b)
    Same as in a), with results from the \nbody\ simulations.}
  \label{fig:period_0_1_fit}
\end{figure}

Figure~\ref{fig:period_0_1_fit}a) shows the resulting initial period distributions as thin solid lines and the final period
distributions as thick solid lines for the process of orbital decay. In
accordance to the results for the semi-major axis distribution the orbital decay (Sec.~\ref{sec:embedded_phase}) reduces the
number of short period binaries. 
In above model the influence of the gas on the binary system has been studied, but the dynamical interactions between the binaries
with other stellar components in the cluster were neglected.  It is known that dynamical  interactions destroy long period binaries.
The question arises whether there is any region in the period distribution which is significantly affected by both processes? 

To investiage this, we want to show the pure dynamical impact of the cluster evolution on the binary population in absence of the ``gas-effect''
described above. So all clusters have been set up without any gas and are initially in virial
equilibrium. Only the first $3~\Myr$ are investigated in this study. As only the most massive stars evolve significantly in this
time, stellar evolution was not considered in the simulations. To improve statistics, 50 realisations of these clusters have been
simulated and their results have been averaged. For more details on the numerical method see \citet{2011A&A...528A.144K}.

To provide comparable initial conditions as in Sec. \ref{sec:embedded_phase} all simulations have been set up as described in
Sec. \ref{sec:clust-binary-popul}. The formal initial binary frequency has been chosen to be 100\% i.e.  each star is generated as
part of a binary system. However, the resulting initial binary frequency is always lower as a consequence of the setup
procedure. All particles in the simulations are created as binaries initially with a semi-major axis sampled from the log-uniform
distribution. Afterwards these binaries are placed randomly in the cluster independent of the semi-major axis. Therefore, binaries
with separations longer than the local mean separation will  often have a closer partner already, so that thereby are fewer large-separation binaries than expected form the set up.

Figure~\ref{fig:period_0_1_fit}b) shows the period distribution of the
simulated cluster initially (thin solid line) and at the age of 3 \Myr\ (solid line) with
the Gaussian fit (dotted line) to the observations. It can be seen that
dynamical evolution of the binary population in a star cluster mainly destroys
wide binaries, but does not affect tight binaries. 
These wide binaries are destroyed in three- and four-body encounters where a
perturber transfers some of it's kinetic energy to the binary system so that
after the interaction, the binding energy of the binary becomes positive
\citep{1975MNRAS.173..729H}. However, this process is only effective if the
binary is not too strongly bound which implies that predominantly wide
binaries are affected. Nevertheless, after 3 Myr still a small amount of wide
binaries exists. Our results give no definite answer to the origin of these
wide binaries. They are either remainders of the initial binary population or
formed by dynamical processes. The reason is that also we perform a large
number of simulations the results for the wide binaries still suffer from low
number statistics.

Similar results were obtained by e.g. \citet{1995MNRAS.277.1491K,2001MNRAS.321..699K} 
using a initial period distributions with a higher number of initial wide binaries 
than used here (dashed line in Fig. \ref{fig:possible_period_distributions}). 
It should be noted that, using a log-normal distribution with a initial lower 
number of wide binaries (thick grey line in Fig. \ref{fig:possible_period_distributions})
\citet{2009MNRAS.397.1577P} obtained a deficiency of wide
binaries in comparison to the field. If the initial binary distribution would be 
log-normal, this would require wide binaries to be formed by other processes see
for example \citet{2010MNRAS.404.1835K,2010MNRAS.404..721M}. However, apart from 
the wide binaries, the overall result does not seem to be
very sensitive to the initial distribution. In our work a log-uniform period
distribution as for the investigation of the orbital decay is applied. 
Otherwise the treatment is very similar to previous work.   

The destruction of the wide binary systems is accompanied by a reduction of
the binary frequency by $13\% \pm 2\%$ after $3~\Myr$ where the error is given
by the standard deviation among the 50 realisations. However, this represents
only a lower limit for the destruction of binaries by dynamical
interactions. As has been mentioned before, although all stars are intended to
be part of a binary initially, a certain fraction of all binaries - dominantly the
  wide - is 'disrupted' during the setup process. Treating these as
being destroyed by the cluster dynamics results in an upper limit of $26\% \pm
2\%$.  In summary, the effect of the dynamical destruction of wide binaries is
a reduction of $\approx 13\%-26\%$ of the binary frequency during the first
3\Myr.  Comparing this to the $12\%$ loss of binaries by merger caused by
orbital decay, this means that both processes are of equal importance, at
least for the considered case.

\begin{figure}
  \centering
  \includegraphics[keepaspectratio,width=0.9\linewidth]{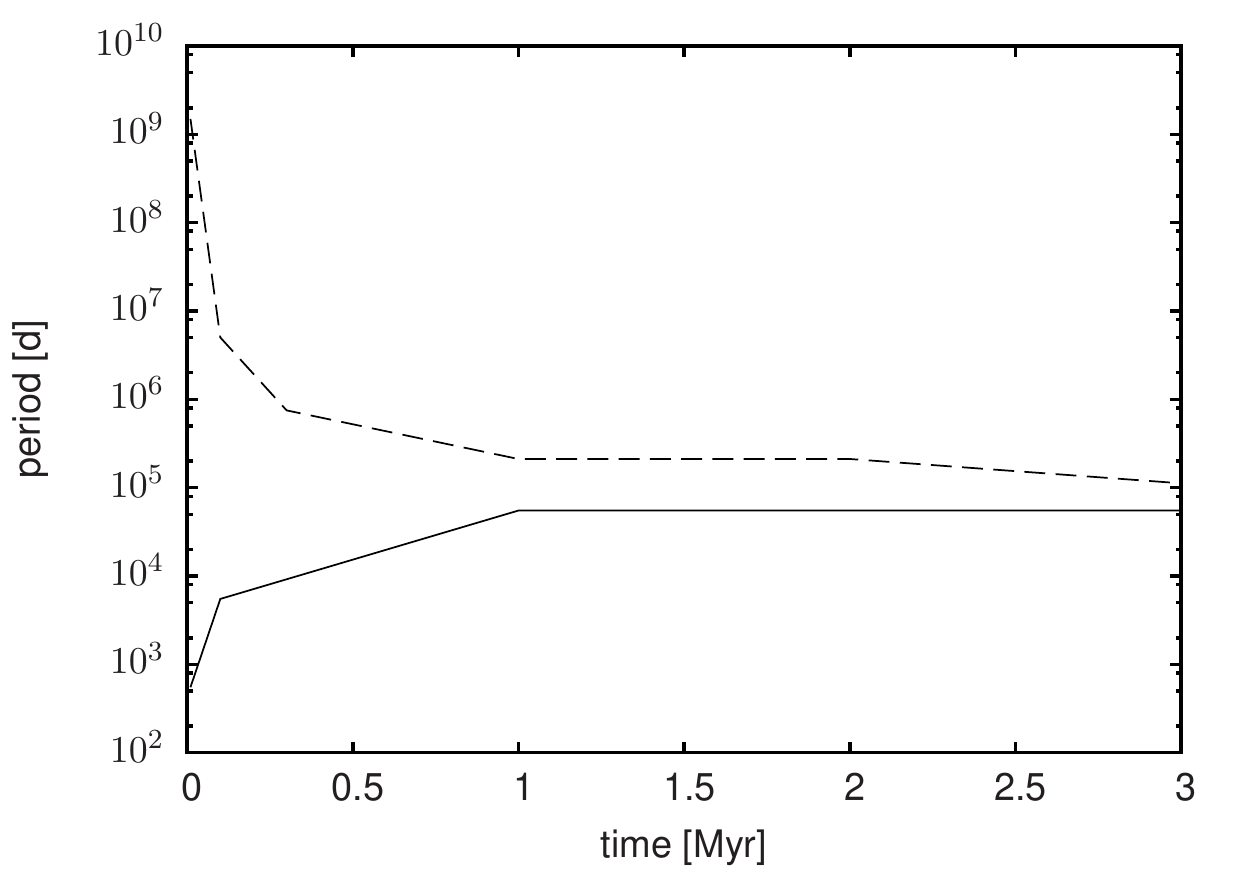}
  \caption{Evolution of $P_{\text{dyn}}$ (upper line) and $P_{\text{orb}}$ in the first $3~\Myr$ of cluster evolution.}
  \label{fig:evolution_effective_ranges}
\end{figure}

In the following we specify the maximum effective period for the orbital decay $P_{\text{orb}}$ and the minimum effective period for
the dynamical destruction $P_{\text{dyn}}$ at which at least 3\% of the binaries are affected by the corresponding processes (see
Fig. \ref{fig:evolution_effective_ranges}). The resulting values are $P_{\text{orb}} = 5.5 \times 10^4$ days and $P_{\text{dyn}} =
1.1 \times 10^5$ days, it follows $P_{\text{orb}} < P_{\text{dyn}}$. So there
is no period range where both processes simultaneously play a role.
Separating the two process as done here, does not take into account binaries
that exchange partners and become more strongly bound. However, testing 
this for the here considered first $3~\Myr$ of the cluster development, we 
found that only in $\ll$ 1\% of all cases
binary harding leads to a transgression into the regime where orbital decay
takes place. This means that both processes can be treated separately (see
Fig.~\ref{fig:evolution_effective_ranges}) in a additive way.  

In order to determine the likelihood that our simulated period distributions
correspond to the observed field population we performed
$\chi^{2}$-tests  as described by \citet{Press:2007:NRE:1403886}:

\begin{equation}
  \label{eq:chi-square-statistic_unequal}
  \chi^{2} = \sum\limits_{i = 1}^{N} \frac{\left(  R_{i} - S_{i} \right)^{2}}{R_{i} + S_{i}}
\end{equation}

where $N$ is the number of bins, $R_{i}$ and $S_{i}$ are the two distributions to be tested. This is only applicable if
$\sum\limits_{i}R_{i} = \sum\limits_{i}S_{i}$. To warrant this we scaled our probability distributions to the total number of
binaries observed by \citet{2010ApJS..190....1R}.  Afterwards the $\chi^{2}$-probability function $Q(\chi^{2}|\nu)$ is used to
calculate the probability that both distributions have been sampled from the same underlying distribution.

\begin{figure}
  \centering
  \includegraphics[keepaspectratio,width=0.9\linewidth]{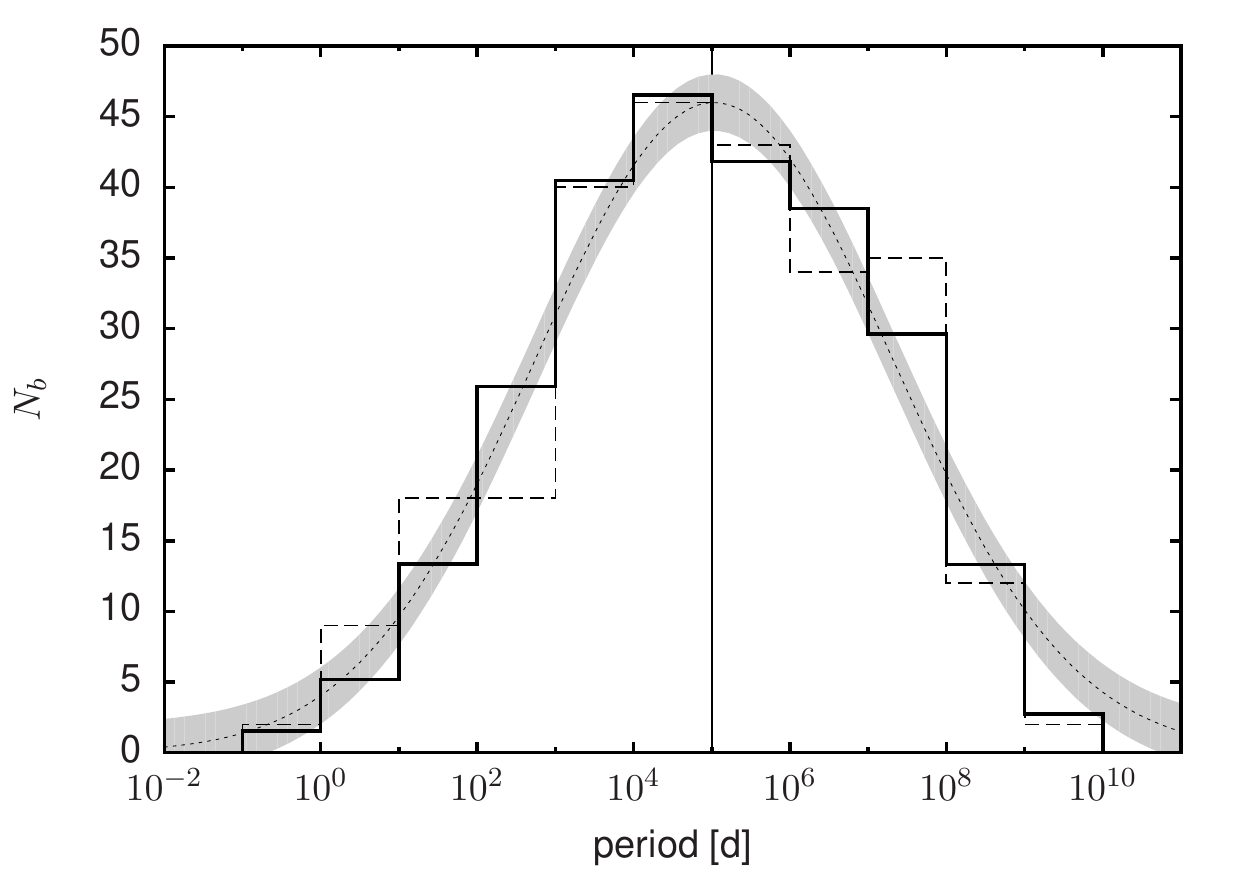}
  \caption{Comparison of the period distributions resulting from the orbital decay of embedded binaries and the dynamical
    destruction (thick solid line) with the observations by \citet{2010ApJS..190....1R} (dashed line). Additionally, the
    log-normal fit by \citet{2010ApJS..190....1R} is shown as short-dashed line.}
  \label{fig:combination}
\end{figure}

Table \ref{tab:chi_square_test_results} shows the results of $\chi^{2}$-tests for the \citet{2010ApJS..190....1R} period
distribution and the simulated one. Obviously, the probability that the initial period distribution and the observed period
distribution of the field have been sampled from the same underlying distribution is negligible (probabilities of $ 3.5 \times
10^{-7} $ and $4.0 \times 10^{-3}$). Similarly, considering only one of the two processes ''orbital decay'' or ''dynamical
interaction'' alone results in (higher, but still) low probabilities. Restricting the $\chi^{2}$-test to the proper period 
ranges ($10^{-1} - 10^{5}$\days\ for the orbital decay and $10^{5} - 10^{10}\days$ for the cluster dynamics) yields much 
better results - both the orbital decay and the cluster dynamics result in $Q(\chi^{2}|\nu)$ values of $67\%$ and $92\%$, 
respectively, indicating already the really good agreement of our simulated period distributions and the observations by 
\citet{2010ApJS..190....1R}. A superposition of the distribution of periods $P< 10^5\days$ depleted by orbital decay and 
the distribution of periods $P > 10^5\days$ resulting from the dynamical destruction is shown as thick solid line in Fig. \ref{fig:combination}. 
Performing the $\chi^{2}$-test on this complete distribution over the entire period range results in a $Q(\chi^{2}|\nu)$-value 
of $94.1\%$.  This clearly demonstrates, that together these two processes naturally reshape a log-uniform distribution to a 
log-normal distribution as observed in the field today.

\begin{table}[h]
  \centering
  \begin{tabular}[h]{c||cc|cr}
    \hline
     & \multicolumn{2}{c}{orbital decay} & \multicolumn{2}{c}{N-body dynamics}\\
    \hline
     & \multicolumn{4}{c}{complete period range}\\                              
     & $\chi^{2}$ & $Q(\chi^{2}|\nu)$    & $\chi^{2}$ & $Q(\chi^{2}|\nu)$ \\
    initial  & 71.5 & $2.3\times 10^{-11}$ & 23.6 & 0.009 \\ 
    final    & 36.4 & $7.2\times 10^{-5}$ & 22.4 & 0.013 \\
    \hline
    \\[1pt]
    \hline
    & \multicolumn{4}{c}{adopted period ranges}\\                          
    & $\chi^{2}$ & $Q(\chi^{2}|\nu)$    & $\chi^{2}$ & $Q(\chi^{2}|\nu)$ \\
    initial & 45.7 & $1.1\times 10^{-8}$  &  6.6 & 0.171 \\
    final   & 3.2  & 0.67                 &  0.9 & 0.92 \\
    \hline
    \end{tabular}
    \caption{$\chi^2$-test results calculated for the initial and final theoretical period distributions and those observed by
      \citet{2010ApJS..190....1R}. The first four column block shows the results of tests against the complete \citet{2010ApJS..190....1R}
      period distribution ranging from $\times 10^{-1} - \times 10^{10}$ \days. The second four column block shows tests
      against the left part of the \citet{2010ApJS..190....1R} period distribution ($\times 10^{-1} - \times 10^{5}\days$) for the
      orbital decay distributions and the right part of the \citet{2010ApJS..190....1R} period distribution ($\times 10^{-5} - \times 10^{10}\days$)}
  \label{tab:chi_square_test_results}
\end{table}

%%% Local Variables: 
%%% mode: latex
%%% TeX-master: "main"
%%% End: 

%% file: conclusions.tex
In this paper we demonstrated for the first time that gas-induced orbital decay of binaries in the embedded cluster phase significantly changes the
properties of short-period binary distribution.  It turns out, this process is of equal importance for the development of the initial binary
 population as the well-studied process of dynamical evolution, at least for ONC-like clusters.

Gas-induced orbital decay not only changes the binary properties, but can even cause mergers of binaries, creating a more massive single star. The likelihood of such mergers depends on
the density profile of the gas and on the separation and the mass of the binary system. For example, while binaries with semi-major axis of $10\,\AU$
rarely merge after $1~\Myr$ in a cluster with a maximum gas density of $10^5 \mathrm{cm}^{-3}$, $40\%$ of all binaries with a semi-major axis of
$0.1\,\AU$ do so. In general the separation between binary systems with $M_{sys} > 10 \Msun$ shrinks on average to less than half its initial value
after $1\,\Myr$.

\begin{figure}
  \centering
  \includegraphics[keepaspectratio,width=0.9\linewidth]{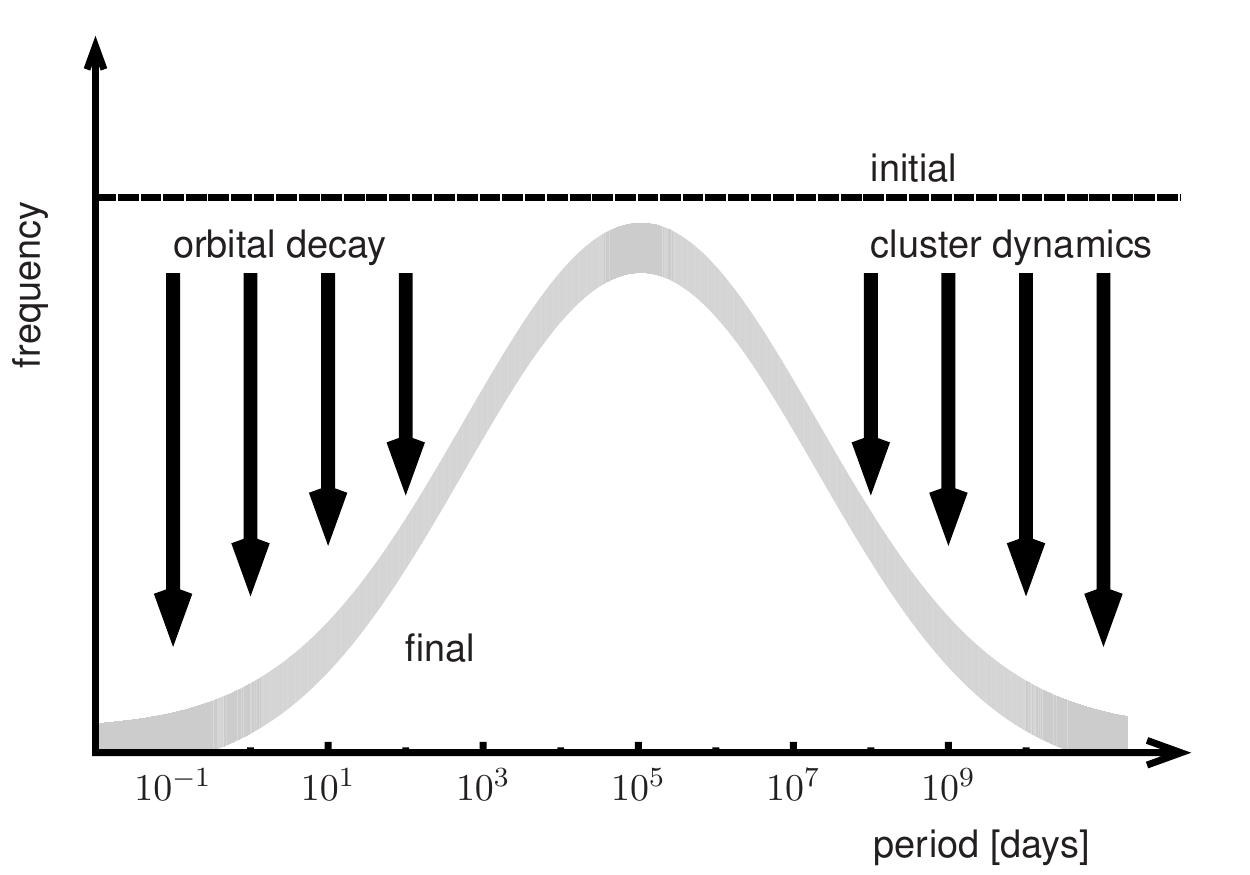}
  \caption{Schematic illustration of the two-stage process forming the log-normal field-binary period distribution observed
    today (grey line) from an initial log-uniform distribution (solid line).}
  \label{fig:sketch_two_stage}
\end{figure}

Orbital decay changes only the period distribution of short period
binaries.  However, long-period binaries, are affected by dynamical
destruction caused by the interaction of the stars in a dense cluster
environment. Our simulations show that
intermediate period binaries are nearly unaffected by either of these two
 processes (see Fig.~\ref{fig:sketch_two_stage}). 

For an ONC-like cluster we found that for these G-type primary stars that the orbital decay due to the interaction of the binaries with the ambient gas basically  only
affects binaries  with periods smaller than $5\times 10^4 \days$  corresponding to separations closer than $\approx 36\AU$. By contrast, dynamical
interactions destroy only binaries with periods larger than $10^5\days$ (i.e. $\approx 53\AU$). This means that there is no
region in the period distribution that is influenced by both effects. Therefore it is possible to investigate the two processes
separately and combine the results.

Perhaps the most striking result of this investigation is, that these two processes transform a primordial log-uniform into a log-normal period
  distribution, on a realistic time scale. The distribution resembles, in all its properties, remarkably that observed in the field, without need of
  further assumptions. Performing a $\chi^{2}$-test of the resulting distribution and recent observations of the field binary population
  \citep{2010ApJS..190....1R} yields a probability of $94.1\%$ that both distributions originate from the same origin.

The emerging picture can be summarized the following way (illustrated in
Fig. \ref{fig:sketch_two_stage}): Binaries born in dense, embedded clusters
 with initially log-uniformly period distributions are processed in two ways. As long as the
cluster is embedded in its natal gas (about 1\Myr), the orbital decay of the
embedded binaries depopulates the left hand side of the period
distribution. The dynamical evolution of the cluster destroys wide binaries,
depopulating the right hand side of the period distribution. The combined
effect of these  equally important processes is that the final period
distribution of the binary population in the star cluster has become
log-normal although it initially has been log-uniform. 

This re-shaping of the period distribution leads to a reduction of the binary frequency by $29\%$.  The gas-induced orbital decay
shrinks the separation between two stars such that mergers in $14\%$ of all cases are possible. The dynamical interactions destroy
$15\%$ of G-type binaries by three-body encounters. This latter fraction only represents a lower limit to the reduction of the
binary frequency due to out specific setup procedure and can be as high as $28\%$. Consequently the overall reduction of the
binary frequency could be up to $42\%$. Assuming that most stars form in an ONC-like cluster and adopting a field binary frequency
of G-type primaries is $58\%$ \citep{2010ApJS..190....1R} our results strongly favour a much higher initial binary frequency of
$87\%-100\%$ (assuming $44\%$ one gets $73-86\%$) for solar-type stars.

Mass segregation and high densities in the cluster centre favour the merging of massive stars. From the merging of predominantly massive stars,
  one would expect to see a difference in the IMF of single and binary stars, with the single star distribution having an excess of massive
  stars. This has not been observed. However, the situation is more complex. Massive stars are as well the most likely to capture a new
  partner. For example mergers which form a stars with $ M > 20 M_{\sun}$, can happen on a timescales of $~ 10^4 yr$
  \citep{2007A&A...475..875P}. The merger product and its companion would become part of the binary population again. So the IMF of the 
  single and binary population would be changed by merger processes. Weather they would be effected to the same degree would require further studies,
  but given the poor statistics at the high mass end, they would be indistinguishable.

Clearly, the field distribution is a mixture of stars originating from stars that experienced star formation in isolation, sparce (Taurus-like), dense
  (ONC-like) and very dense (Arches-like) clusters. On the one hand,  associations like the Taurus clusters probably never had a gas density above $10^4 cm{-3}$ and the current
  stellar density is below 10 stars $pc^{-3}$ \citep{2009ApJ...703..399L}. This means in such systems the binary population will neither be affected
  by orbital decay nor orbital destruction in a significant way and is therefore nearly unaltered from its primordial state. On the other hand, orbital decay becomes important for high gas densities and dynamical destruction for high stellar densities

On the basis of actual observations, one can only speculate how many stars are born in these different density regions. While \citet{2010arXiv1009.1150B}
stated, that only a minor part of all stars in the
solar neighbourhood form in high density regions, this is unclear for the
total of the galaxy. \citet{2011arXiv1111.3693D} concluded, that 1/2 - 2/3  of all
stars are in clusters with more than  1000 stars. As such massive clusters
initially had a much higher stellar densities \citep{2009A&A...498L..37P}, this indicates
that environmental effects are important for the latter. The remarkable resemblance of the distribution, obtained after both processes
have taken place, to the log-normal period distribution of the field binary
population is a strong indication, that ONC-like clusters might be a dominant
contributor to the field distribution. 

Additional to the here studied ONC-type clusters, further investigations should
include a variety of  initial conditions: different stellar and gas density
distributions \citep{1995MNRAS.277.1522K,2011arXiv1108.3566P}, a range of cluster
densities \citep{2010A&A...509A..63O,2011arXiv1106.5050M}  and different virial
states of the cluster (Allison 2009). Currently we test the limitations of the
model of orbital decay by numerical simulations  (Korntreff $\&$ Pfalzner, in prep). 

To verify our initial conditions observationally, detailed studies of binary populations in very young ($< 100 000$ yrs) embedded star clusters,
  before the onset of the here investigated processes, would be necessary.  However conclusions about the binary distribution in such young systems
  would be hindered by low-number statistics. Usually only a few tens of stars are observable, although the true membership might be considerably
  higher due to the high extinction in these clusters. Combining data from different such clusters would be an alternative. An other approach could be
  to observe different regions within a single OB associations. The processes are predominant at work in the association centre and much less so in
  the outskirts. Therefore, it could be expected that the period distribution in the association centre differs considerably from the outskirts -
  close to the log-normal in the cluster center and log-uniform in the outskirts. However, this picture neglects the mixing within the cluster, so
  that further theoretical work is needed to investigate this point.

%Both processes inevitably take place in young dense star
%clusters and mainly depend on the actual gas and stellar densities.  

%% Both processes inevitably take place in young dense star clusters and mainly depend on the actual gas and stellar densities. This
%% allows us the following strong conclusion: Because the form of the period distribution of the field binary population and that
%% evolved in our ONC model cluster are the same, the star cluster environment where most stars formed are likely to have had a similar
%% stellar density to that of the ONC. Similarly the gas density of the common star forming environment must be somewhere around
%% $10^5\,\mathrm{cm}^{-3}$. Our results thus provide independent evidence that the ONC is a typical star-forming region. Hence, our
%% results are potentially applicable to most star forming region. So the properties of today's field binary distribution give direct
%% access to the conditions under which star formation takes place.

%%% Local Variables: 
%%% mode: latex
%%% TeX-master: "main"
%%% End: 